\documentclass[aps,prb,showpacs,twocolumn,groupedaddress]{revtex4}

\begin{document}

\title{Pretransitional phenomena in dilute crystals
with first-order phase transition}

\author{P. N. Timonin}
\email{timonin@aaanet.ru}
\affiliation{Physics Research Institute at Rostov State University
344090, Rostov - on - Don, Russia}

\date{\today}

\begin{abstract}
Pretransitional phenomena at first-order phase transition in crystals
diluted by 'neutral' impurities (analogue of nonmagnetic atoms in dilute
magnets) are considered. It is shown that field dependence of order parameter becomes nonanalytical in the stability region of the ordered phase, while smeared jumps of thermodynamic parameters and anomalous (non-exponential) relaxation appear near transition temperature of pure crystal.
\end{abstract}


\maketitle
In last decades the intensive theoretical and experimental studies
of phase transitions in disordered crystals reveal the drastic influence which
frozen impurities and defects may have on the transition.
They can also give rise to a variety of pretransitional phenomena.
Such phenomena were first considered by Griffiths \cite{1}
in the framework of dilute Ising model, which has empty sites (without spins)
randomly scattered over lattice. He had shown that in such dilute magnet
the transition is preceded  by specific phase, where magnetization is
nonanalytical function of magnetic field.

While the bulk transition in this model takes place in the (infinite)
percolation cluster of spins, the nature of  pretransitional effects lies
in the properties of finite magnetic clusters. Due to the short-range
(nearest neighbor) interaction between spins all magnetic clusters are
completely independent and give additive contributions to the
thermodynamic parameters of the system. Immediately below the
transition point of pure magnet large clusters begin to develop high
thermodynamic barriers between the states with all spins up or down.
With due reservations these processes can be called 'local phase transitions'.
As there is a finite probability to find arbitrarily large magnetic clusters
with arbitrarily high barriers this results in nonanalytical field
dependence of magnetization \cite{2,3,4}. Also same barriers give
rise to anomalously slow nonexponential relaxation of the order parameter
in this Griffiths' phase \cite{5}.

Apparently, the existence of Griffiths' phase is not limited to the case
of dilute Ising magnet. It should be also present in all crystalline solid
solutions near generic second-order phase transition with short-range
order parameter interaction, if it takes place in only one component of
the solution. The other 'neutral' component(s) of the solution (impurities)
is supposed to have no influence on the transition in the 'pure' copmponent
being an analogue of nonmagnetic impurities or empty sites in the dilute 
Ising model. Hereafter we refer to this generalization of the notion of
dilute magnet as 'dilute crystal with a (second-order) transition' or just
'dilute crystal'.

While the Griffiths' anomalies near second-order phase
transitions are thoroughly investigated, now nothing is known about
the pretransitional phenomena near first-order transitions in such dilute
crystals, in spite of the attention paid to a smearing of these transitions
by frozen disorder\cite{6,7,8}. But it is evident that some similar anomalies
could exist in this case, as high barriers between ordered and disordered
states of finite clusters of pure component also appear above the transition
point of pure percolation cluster (if it exists).

Let us consider the first-order transition with one-component order parameter 
$\varphi$, described in a pure crystal by the inequilibrium thermodynamic potential, which is even function of $\varphi$. We assume that the dimensionless (divided by the temperature $T$) density of such potential $ f\left( \varphi  \right) $ has a minimum at $ \varphi = 0$ for $T>0$ and two another minima at$\varphi \pm \varphi_s$ when $T<T_+$, $T_-<T_+$. Below some $T_0$, $T_-<T_0<T_+$, $f_s \equiv f \left(\pm \varphi_s \right)$ becomes less than $f_0 \equiv f \left( 0 \right)$, so the first-order transition takes place at $T_0$ in pure crystal.

It can be shown that in dilute crystal the dependence of the order parameter on the conjugate field becomes nonanalitycal below $T_+$. 
To do this we consider the contributions to the average order parameter from large compact clusters of pure component with a number of sites much greater than that of neighboring impurities and diameter much larger than order parameter correlation length. Such clusters can be described by the density of (inequilibrium) thermodynamic potential of infinite pure crystal. Their order parameter distribution in the external field $h$ has the form
\begin{equation}
\rho _n \left( \varphi,h  \right) = Z_n^{ - 1} \exp
\left[ { - nf\left( \varphi  \right)}-nh\varphi \right]\label{eq:1}
\end{equation}
\[
Z_n  = \int {d\varphi } \exp \left[ { - nf\left( \varphi -nh\varphi \right)}
\right]
\]
where $n$ is number of sites in a cluster. It is easy to obtain the average order parameter for a cluster
\[
\left\langle \varphi  \right\rangle _{T,n}  \equiv \int {d\varphi \varphi
 \rho _n \left( \varphi, h  \right)},
\]
in a small field 
\[
\left|h\right| \le \varphi_s/\max\left(\chi_0, \chi_s \right),
\]
\[
\chi_0 = 1/f''\left(0\right),\\
\qquad
\chi_s = 1/f''\left(\varphi_s\right).
\]

For $n >> 1$ at $T_-<T<T_+$ we have
\begin{equation} 
\left\langle {\varphi \left( h \right)} \right\rangle _{T,n}  = \frac{{\chi _0 h + 2\varphi _s \left(h\right) \left( {\chi _s /\chi _0 } \right)^{1/2} e^{ - n\delta f} \sinh \left( {nh\varphi _s } \right)}}{{1 + 2\left( {\chi _s /\chi _0 } \right)^{1/2} e^{ - n\delta f} \cosh \left( {nh\varphi _s } \right)}} \label{eq:2}
\end{equation}
while at $T<T_-$ 
\begin{equation}
\left\langle {\varphi \left( h \right)} \right\rangle _{T,n}=\varphi_s \left( h \right).
\label{eq:3}
\end{equation}

Here $\delta f = f_s - f_0$ and 
\[
\varphi_s \left( h \right)=\varphi_s\tanh \left( {nh\varphi _s } \right) +
\chi_s h. 
\]

Using the numbers of compact clusters per site, $N (n)$, one can find
the average order parameter
\[
\left\langle \varphi  \right\rangle  = \sum\limits_{n > n_0 }
{N\left( n \right)} \left\langle \varphi  \right\rangle _{T,n}.
\]
Here sum is taken over sufficiently large $ n > n_0$ for which Eqs. (\ref{eq:2}) and (\ref{eq:3})
 is valid,
\[
n_0 = c\xi_{max}^d \gg 1,
\]	
$\xi_{max}$ being the largest of correlation lengths of two phases
(in lattice units) when $T_- < T < T_+$, $c \gg 1$ and $d$ is space dimension.

When $\delta f < 0$ the expression in Eq. (\ref{eq:2}) has poles in the complex $h$ plane which are close to that of $\tanh \left( {nh\varphi _s } \right)$ 
\[
h_{n,m}  \approx  \pm \left( {4m + 1}\right)\frac{{i\pi }}{{2n\varphi _s }},~  m = 0,1,...
\]
They are arbitrarily close to $h  = 0 $ for large $n$. As $N \left( n \right) \ne 0$ at all $n$ the expansion of average order parameter in powers of $h$ would not converge at $T < T_0$.

Above $T_0$ the expression in Eq. (\ref{eq:2})has poles at
\[
h_{n,m}  \approx  \pm \delta f/\varphi \pm i\left( {2m + 1}\right)\frac{{2\pi }}{{n\varphi _s }},~  m = 0,1,...
\]
which are close to $\pm h_c  = \pm \delta f/\varphi _s $
for large $n$. So at $T>T_0$ average order parameter $\left\langle {\varphi \left( h \right)} \right\rangle$ is nonanalytical at the fields $h=h_c$ and $h=-h_c$ in which transition in pure crystal takes place. We can also expect that the contributions to $\left\langle {\varphi \left( h \right)} \right\rangle$ from sparse clusters would have the similar property at larger $h$ as they have, presumably, smaller $\varphi _s $ and larger $\left|\delta f \right|$ at the same $T$. Then the field dependence of the average order parameter will be nonanalytical at all $\left|h \right| > h_c$. 

The 'local transitions' in finite clusters also manifest themselves in the temperature dependence of thermodynamic parameters at $h=0$. Consider the compact cluster contribution to the zero field susceptibility
%\begin{eqnarray*}
\[
\left\langle \chi  \right\rangle  = \sum\limits_{n > n_0 }{N\left( n \right) \left\langle  \chi  \right\rangle _{T,n}}
\]
\[
\left\langle \chi  \right\rangle_{T,n} \equiv
n\left[ \left\langle {\varphi ^2 } \right\rangle _{T,n}
- \left\langle \varphi \right\rangle _{T,n}^2 \right]
\]
%\end{eqnarray*}
At $T_-<T<T_+$ we have
\begin{equation}
\left\langle \chi  \right\rangle_{T,n} \approx \frac{{\chi _0^{3/2}e^{n\delta f}+2\chi _s^{3/2}+2n^2\varphi_s^2\chi _s^{1/2} }} {{\chi _0^{1/2} e^{n\delta f}
+ 2\chi _s^{1/2} }}, \label{eq:4}
\end{equation}

$N (n) $ is given by the general formulae \cite{9},
\[
N(n) = \sum\limits_s {g(n,s)p^n \left( {1 - p} \right)^s } {\rm  }
\]
where $1- p$ is the concentration of impurities, and $g(n,s)$
is the number of pure clusters with $n$ sites and $s$ neighboring impurities.
The compactness of clusters considered can be accounted for by restriction
of the sum to $s \ll n$. In the following we will use the lower estimate for $N (n)$,
\begin{equation}
N(n) \approx p^n \left( {1 - p} \right)^{s_{\min } \left( {n,d} \right)}
{\rm }\label{eq:5}
\end{equation}
\[
s_{\min } \left( {n,d} \right) = e_d n^{\left( {d - 1} \right)/d},
~e_d {\rm  } = {\rm   }\left[ {2\pi ^{d/2} {\rm  }d^{d - 1} /
\Gamma (d/2)} \right]^{1/d}
\]
with $s_{\min } \left( {n,d} \right)$ being the area of
$d$ - dimensional sphere, which bounds the volume $n$. According to 
Ref. (\onlinecite{10}) Eq. (\ref{eq:5}) describe correctly the main features 
of large $n$ behavior of $N (n)$ either for $p \to 1$ or $p \to 0$ except for
a change of $p$ to some $p' > p$ when $p \to 0$ (see Ref. (\onlinecite{4})).

Near $T_0$, when the condition
\[
n_0^{-1} \ll \left| {\delta f} \right| \ll \min \left\{ {\ln \left( {\frac{1}{p}}  \right),e_d^{d/\left( {d - 1} \right)} \ln ^{d/\left( {d - 1} \right)} \left( {\frac{1}{{1 - p}}} \right)} \right\}
\]
can be fulfilled, we get from Eqs. (\ref{eq:4}), (\ref{eq:5})
\begin{eqnarray*}
 \left\langle \chi  \right\rangle  = \left\{ \begin{array}{l}
 \nu _0 \chi _0 ,~ T > T_0  + \delta T \\ 
 \nu _0 \chi _s  + \nu _2 \varphi _s^2 ,~ T < T_0  - \delta T \\ 
 \end{array} \right. \\ 
 \nu _k  \equiv \sum\limits_{n > n_0 } {n^k N\left( n \right)}  \\ 
 \end{eqnarray*}
Here $\delta T$ is defined by the condition $n_0 \left|\delta f \right| \gg 1$, so
\[
\delta T \sim T_0 /n_0 \delta S,
\]
where $\delta S$ is entropy jump at the transition in pure crystal.
Thus near transition point of pure crystal there is slightly smeared
jump in the zero field susceptibility.

Also near $T_0$ the average entropy of large compact clusters,
\[
\left\langle S \right\rangle  = \frac{\partial }{{\partial T}}
\sum\limits_{n > n_0 } {N\left( n \right)}T\ln Z_n ,
\]
mimics the behavior of pure crystal entropy with prefactor $\nu_0$.

Let us also consider the average modulus of the order parameter, $\left\langle {\left| \varphi  \right|} \right\rangle $. This quantity can be interpreted as measure of the local order and could be observed in local experiments such as NMR. In pure (infinite) crystal it becomes nonzero below $T_0$ simultaneously with $\left\langle \varphi  \right\rangle $. In dilute crystal $\left\langle {\left| \varphi  \right|} \right\rangle \ne 0$ at all $T$. Indeed, for the compact clusters we have
\begin{eqnarray*}
\left\langle {\left| \varphi  \right|} \right\rangle_{T,n} \approx \left(8\chi_0/\pi n\right)^{1/2},~T>T_+\\
\left\langle {\left| \varphi  \right|} \right\rangle_{T,n} \approx \varphi_s,~T<T_-
\end{eqnarray*}
while at $T_-<T<T_+$
\begin{equation}
\left\langle {\left| \varphi  \right|} \right\rangle _{T,n}  = \frac{{\left( {8\chi _0 /\pi n} \right)^{1/2}  + 2\varphi _s \left( {\chi _s /\chi _0 } \right)^{1/2} e^{ - n\delta f} }}{{1 + 2\left( {\chi _s /\chi _0 } \right)^{1/2} e^{ - n\delta f} }}\label{eq:6}
\end{equation} 

Thus in dilute crystal $\left\langle {\left| \varphi  \right|} \right\rangle = \sum\limits_{n > n_0 } {N\left( n \right) \left\langle {\left| \varphi  \right|} \right\rangle _{T,n}}$ has small nonzero value at all temperatures above the transition point. 

We cannot describe in the same manner the (additive) contributions of the
sparse clusters with $s \ge n$ to the thermodynamic parameters. Yet we may
suppose that they undergo similar 'local transitions', but at temperatures
lower than $T_0$ due to the lower average coordination number. Then
continuous sequence of these transitions below $T_0$ will result in steep
increase of $\left\langle \left| \varphi \right| \right\rangle$ and decrease of
$\left\langle S \right\rangle$ at $T <  T_0$ for all concentrations $p$,
 while the bulk transition takes place at lower temperature (less than $pT_0$
\cite{11}) or vanishes altogether below percolation threshold.

We can also show that in the region where phases coexist large compact clusters
have anomalously slow dynamics with broad spectrum of relaxation times. Here 
the most slow relaxation process is the decay of metastable state. In case of 
one-component nonconserved order parameter it normally proceeds via
nucleation and growth of droplets of stable phase \cite{12,13}.
As the nucleation barrier must be overcome the time it takes is
\[
\tau _c  = \tau _1 \exp \Delta,
\]
\[
\Delta  = \mathop {\max }\limits_l \left( {dl^{d - 1}
\sigma  - l^d \left| {\delta f} \right|} \right) = \sigma l_c^{d - 1},
\]
Here  $\sigma$ is dimensionless interphase surface tension
(divided by $T$ and multiplied by the area of unit cell face),
$l_c$ is the critical droplet size,
\[
l_c  = \left( d - 1 \right)\sigma /\left| {\delta f} \right|
\]
 and $\tau_1$ is the time inverse proportional to
the droplet growth rate \cite{12}.

The decay time $\tau_c$ does not depend on system size, which is implicitly
assumed to be much greater than $l_c$. Yet for finite clusters with size
less than that of critical droplet, $n < n_c \equiv l_c^d $,
the above relaxation process transforms into the sweeping of domain
wall through a cluster. Then size-dependent barrier,
\begin{equation}
\Delta _n  = n^{\left( {d - 1} \right)/d}
\sigma  - n\left| {\delta f} \right|/2,\label{eq:7}
\end{equation}	
should be overcome, which takes the time
\begin{equation}
\tau _n  = \tau _1 \exp \Delta _n.\label{eq:8}
\end{equation}	

Far from $T_0$ most clusters exhibit the exponential long-time relaxation
with single relaxation time $\tau_c$.  When $T \to T_0$ , $l_c \to \infty$
and the large part of clusters have size-dependent relaxation times.
Thus in dilute crystal near $T_0$ the broad spectrum of relaxation times
emerges. To describe this quantitatively let us consider a contribution of
compact clusters to a long-time asymptotics of average dynamic correlator
\[
\delta \left\langle {C\left( t \right)} \right\rangle
 = \sum\limits_{n_0  < n}{N\left( n \right)C_n \left( t \right)},
\]
\[
C_n \left( t \right) = \left\langle {\varphi \left( t \right)\varphi
\left( 0 \right)}\right\rangle _{T,n}  - \left\langle \varphi
 \right\rangle _{T,n}^2 .
\]
Here $C_n \left( t \right)$
describes relaxation in $n$-site compact cluster and according to above
consideration we have for $t \to \infty$
\[
C_n \left( t \right) = \left\{ \begin{array}{l}
 A_n \exp \left( { - t/\tau _c } \right),{\rm   }n > n_c  \\
 A_n \exp \left( { - t/\tau _n } \right),{\rm   }n < n_c  \\
 \end{array} \right.
\]

 Thus near $T_0$ when $n_c  > n_0 $
\[
 \delta \left\langle {C\left( t \right)} \right\rangle  =
 \left\langle A \right\rangle \exp \left( { - t/\tau _c }
 \right) + B\left( t \right)
\]
\[
 \left\langle A \right\rangle  = \sum\limits_{n_c  < n}
 {N\left( n \right)A_n }
\]
\begin{equation}
B\left( t \right) = \sum\limits_{n_0  < n < n_c }
{N\left( n \right)A_n \exp \left( { - t/\tau _n }\label{eq:9}
\right)}
\end{equation}

Assuming $A_n$ to be a slow varying function of $n$ (slower than $exp(an)$),
we can use the steepest descend method to find time dependence of $B(t)$
 for large $t$. From Eqs. (\ref{eq:7})-(\ref{eq:9}) it follows
that at $t >> \tau_c$ clusters with
$n_*  = \left( 2/d \right)^d n_c $, having the largest barrier
for a sweeping domain wall, give main contribution to $B(t)$,
\begin{eqnarray*}
B\left( t \right) \sim \exp \left( { - t/\tau '_c } \right),~t >> \tau_c \\
\tau '_c  = \tau _1 \exp \left[ \left(\sigma/d\right)
\left( 2l_c /d \right)^{d - 1}\right].
 \end{eqnarray*}

When
\begin{equation}
\tau _0  < t <  < \tau _c,\label{eq:10}
\end{equation}
\[
\tau _0  \equiv \tau _1 \sigma ^{ - 1}
\exp \left( {\sigma n_0^{\left( {d - 1} \right)/d} } \right)
\]
the main contribution to $B(t)$ give clusters of size
$n_*  = \left[ {\sigma ^{ - 1} \ln \left( {\sigma t/\tau
_1 } \right)} \right]^{d/\left( {d - 1} \right)} $
and we get
\begin{equation}
B\left( t \right) \sim \left( {\frac{{\tau _1 }}{t}} \right)^{\lambda
\left( p \right)} \exp \left\{ {\ln \left( p \right)\left[ {\sigma ^{ - 1}
 \ln \left( {\sigma t/\tau _1 } \right)} \right]^{d/\left( {d - 1}\label{eq:11}
\right)} } \right\})
\end{equation}
\[
\lambda \left( p \right) =  - e_d \sigma ^{ - 1} \ln \left( {1 - p} \right).
\]

Exponential term in (\ref{eq:11}) falls slower than pure exponent but faster than all powers of $1/t$. Yet for $p$ close to 1 its variation can be small as
compared to that of prefactor in all region (\ref{eq:10}) and this
term should be dropped leaving only power-law relaxation.
This could happen when
\[
l_p  \equiv \frac{{\lambda \left( p \right)\sigma }}{{\ln \left( {1/p} \right)}} =
 e_d \frac{{\ln \left({1 - p} \right)}}{{\ln p}} >  > 1
\]

Then, not very close to $T_0$ , for $l_c << l_p$ we have power-law relaxation
in the interval (\ref{eq:10}) described solely by prefactor in (\ref{eq:9}), while in the
immediate vicinity of $T_0$ for $l_c >> l_p$ this prefactor can be dropped
in (\ref{eq:11}).

Thus near $T_0$ there appears time interval (\ref{eq:10}) with non-exponential
relaxation which expands to infinity right at $T_0 $.
Owing to the common relaxation mechanism, same Eq. (\ref{eq:11})
holds for long time asymptotic in original Griffiths' phase near
second-order transition with one-component nonconserved order parameter.
It was first derived in Ref. \onlinecite{5}  without power-law prefactor since
author used the estimate $N\left( n \right) \sim p^n $, which does not hold
for $p$ close to 1.

Sparse pure clusters with $s \ge n$ would also exhibit non-exponential
relaxation of a similar origin. We cannot give an exact description of their
contributions to the dynamic correlator. Yet at every temperature below $T _0$
a 'local transition' takes place in some specific sort of sparse pure clusters and their relaxation would dominate the infinite time asymptotic. So the above results for compact clusters describe the order parameter relaxation at $t \to \infty$ in close vicinity of  $T _0$ only, while far below it relaxation of sparse clusters must be considered.

Here we may also note that quite similar dynamics will appear above the isostructural first-order transitions (of the liquid-vapor type), which is described by the potential with two minima, one at $\varphi = \varphi_0$ and another at $\varphi = \varphi_s$, corresponding to high- and low-temperature phases. Also the smeared jumps of the susceptibility and the entropy appear at $T_0$ in this case, while the field dependence of the average order parameter becomes nonanalytical near $T_0$ at $h_c = \delta f/\left(\varphi_s - \varphi_0 \right)$.

To conclude, the present qualitative considerations show that pretransitional phenomena in dilute crystal with first-order phase transition has features rather similar to that of the original Griffiths' phase. They can be generalized to the first-order transitions with multicomponent order parameters and other types of random media, which also have cluster size distributions spreading up to $n = \infty$. So one may expect that analogous phenomena could be observed, for example, near melting points or the liquid crystal transitions in the substances confined in porous media.

\begin{acknowledgments}
This work was made under support from INTAS, grant 2001-0826.
I gratefully acknowledge helpful and illuminating discussions with
V. P. Sakhnenko.
\end{acknowledgments}

\end{document}